# CARRIER SENSE MULTIPLE ACCESS TUNING PARAMETERS USING GAME THEORY

Mahdieh Ghazvini[1], Naser Movahedinia[1], Kamal Jamshidi[1]

[1]Department of Computer Engineering, Faculty of Engineering, University of Isfahan, Isfahan, Iran

`{ghazvini, naserm, jamshidi}@eng.ui.ac.ir`

## ABSTRACT

*Ad Hoc and Mesh networks are good samples of multi agent systems, where their nodes access the channel through carrier sense multiple access method, while a node channel access influence the access of neighbor nodes to the channel. Hence, game theory is a strong tool for studying this kind of networks. Carrier sense multiple access parameters such as minimum and maximum size of contention window and persistence factor can be modified based on game theoretic methods. In this study different games for tuning the parameters is investigated and different challenges are examined.*

## 1. INTRODUCTION

Medium access layer controls the way of access to common media and permission for transmission over them. Medium access models are divided into two general classes of contention-based and contention-free models. Contention-based models include random access methods. Random access methods are divided into carrier sensing models like CSMA and no carrier sensing models like ALOHA. In CSMA, probability of channel access is implemented through a back off algorithm on contention window (CW) or as persistence probability (P). In back off mechanism, each wireless node keeps a contention window, before transmission, it waits for a random period of time that is limited to the size of contention window. In persistence mechanism, each wireless node keeps a persistence probability and upon observing a channel idle, it accesses the channel with this probability. If persistence mechanism is implemented, the probability of channel access is the persistence probability (p) and in case of using back off mechanism, channel access probability is related to the size of contention window (CW) as follows[1-3]:

$$p = \frac{2}{CW_{\min} + 1} \qquad (1)$$

Choosing the contention criteria and the contention resolution algorithms are performance keys of medium access methods. Improper selection of these parameters leads to poor performance. One of common models of CSMA is DCF algorithm that will be referred to briefly. In DCF, each node to transmit a new packet monitors the channel activity. If the channel is idle for a periods of time, called DCF Inter frame Space (DIFS), the node starts its transmission. Otherwise, the node insists on monitoring the channel so as to see it idle to the amount of DIFS. In the first attempt for transmission, contention window is tuned to the size of minimal contention window ($CW_{min}$). After each unsuccessful transmission, CW will be larger in order to reach its maximal value namely $CW_{max}$. Once CW reaches $CW_{max}$, this value remains constant so that the packet is transmitted successfully or the number of retransmissions reaches retry bound.

In this way retransmission is stopped and the packet is dropped [4]. In the original version of DCF, each new transmission starts with $CW_{min}$ value regardless of the contention level in the network. Therefore, there is a need to other models for determining the size of $CW$ or persistence probability in order to reach high performance (throughput and low collision) and a better fairness.

As nodes decide independently about their channel access and access of a node to the channel affects the neighbouring nodes, game theory is a powerful tool for studying this kind of network [2, 4-7].

## 2. GAME THEORY

Game theory is a branch of applied mathematics which is used in most of sciences. Whenever utility and profit a person is looking for is not only effected by his own attempt and decision but also by attempt and decision-either positive or negative- of another party, it is referred to as a game. Game theory is a tool for analysis of systems as well as a tool for optimizing. Although optimization theory is not able to consider interactions between different players (like users), game theory has turned out to be a good approach for studying and analyzing the behavior of complicated and interacting systems [7,8]. The main feature of decision making in game is that each player before decision making should analyze the reaction of other players toward him and then make a decision that is the best for him and contains the most benefit with respect to reaction of another party. An environment in which exists such effect and mutual reaction among decisions of individuals is called a strategic environment and each decision maker in this environment is called a player[8, 9].

A game consists of a set of players, a set of movements and a set of payoff functions. Usually in every game, a payoff function is defined as the subtraction of the utility function and the user cost. Utility function is a mapping of all behaviours of player into a set of real numbers, its definitions plays an important role in the game. In some researches, however, payoff function and utility have been considered equivalent. Strategy of an agent can be every action from his actions space or a combination of them. If players choose a behaviour clearly, it is called pure strategy but generally players choose their strategy with probability and do not have complete confidence in selection of opposite player; such a behaviour is called mixed strategy. Mathematical representation of a game is $\{N, A_i, \{u_i\}\}$ which represent set of players, action space and set of payoff functions, respectively[10, 11].

Utility is a criterion for measuring a node satisfaction level resulted of its actions. Utility function is a reward (Reward minus cost) that is expected to be obtained by each link from the network after their transmission. Users have different utility functions with respect to the kind of their services. With maximizing the network utility (for example, total utility of all users) social welfare of system is also maximized. On the other hand, utility functions can be interpreted as handles for controlling tradeoff between performance and fairness. In game theory, normally, the goal is finding equilibrium(s) in the game. In equilibrium point, each player chooses a strategy that is the best response to selected strategies of the other players. Several equilibria have been proposed, among which the most well-known is Nash equilibrium. Nash equilibrium is a solution from a game including a number of players in which no user gains benefit through one-way change of her/his strategy [12-16].

Players can be cooperative or non-cooperative. In the first case, players cooperate with each other and the problem is converted to an optimization problem that each player leads the system toward a social equilibrium. In a cooperative game, players through coalition decision making, bargain, dealing, and discussion try to reach agreement so that they may gain maximal benefits. The most standard criterion for expressing equilibrium performance in cooperative games is Pareto performance. Pareto means that a user cannot increase his utility without decreasing the utility of at least one of the other players. Another kind of games is non-cooperative game in which each decision maker chooses his strategy selfishly. In fact, in non-cooperative games each player chooses strategies without coordinating with others. In this case, equilibrium if any will be Nash equilibrium.

## 3. GAME THEORY IN WIRELESS NETWORKS

Game theory studies decision making in a shared environment by some decision makers with different goals. Nodes of a network, is a good example of such a condition. Hence, game theory in wireless networks especially in mesh and Ad Hoc networks is applied and such networks can be easily modeled by game theory. For mapping network components to a game, usually network nodes and all resources required by nodes for communication (spectrum, power, bandwidth, time, etc) are considered as players and resources respectively[9, 10]. In wireless networks games the strategy space can include some actions related to network functions for example, decision on whether transmit the packets, forward or no, tuning of power level, selection of modulation method, contention window adjustment or persistence probability adjustment, regulation of hold off parameter, etc. finally Utility or payoff functions can be defined based on some performance criteria like utilization, fairness, throughput, target SNR as listed in Table1[11, 17-20].

Table1. Mapping Between Wireless Networks Elements and Game Theory Elements

| Components of games | Components of wireless networks |
|---|---|
| Players | Network links, Network nodes |
| Strategy | Transmission power, Bandwidth, transmission time, Transmission probability, Contention window, Forwarding a packet or no. |
| Payoff | Delay, Channel utilization, Fairness, Throughput, Power consumption, Target SNR, Packet loss ratio |

In recent years game theory has been mainly used in some problems like routing and resources allocation in competitive environments. The main advantages of game theory in wireless networks are analysis of distributed systems, cross layer optimization and designing of encouraging methods for avoiding selfish behaviors. Application of game theory, its models and challenges in networks have been well studied in [2, 8-10, 17, 21-26]

## 4. CARRIER SENSE MULTIPLE ACCESS TUNING PARAMETERS BASED ON GAME THEORY

Medium access control is a distributed approach in access to a shared wireless channel among competitive nodes. Since wireless networks use a shared transmission medium, collision may occur because of simultaneous transmissions by two or several interfering nodes. In order to control the contention between transmissions of different nodes, increasing the performance of system and fairness, it seems necessary to use distributed media access protocols.

In random access games, the strategy of a player is usually transmission probability or equivalently, the size of her/his contention window. Payoff functions include utility portion from channel access and packet collision cost [27, 28]. By studying the previous works it was revealed that strategies of more players were contention window, transmission power and data rate. Following, proposed CSMA games, from two perspectives of cooperative and non-cooperative games has been investigated.

It has shown that in Ad Hoc networks in which all nodes cooperate with each other, game theory can be used. In cooperative games, agents interact with each other and choose their strategies after agreement. Usually implementation of cooperative solutions is difficult and requires interaction of agents with each other and needs to be repeated some times. But generally cooperative solutions are much better than non-cooperative solutions.

Incomplete cooperative game theory is possible through estimating the current state of the game (e.g. the number of competing nodes) and achieves equilibrium strategy of each node based on estimation of game state. Zhao et al. have proposed an incomplete cooperative game for improving the performance in Ad Hoc [6] and mesh networks [7]. In these games each node estimates the number of competing nodes (n) in the network using the method presented in [29, 30]. Then it tunes its equilibrium strategy through adjusting its minimum size of contention window (CW$_{min}$) as follows:

$$CW_{min} = \begin{cases} \lfloor n \times rand(6,7) \rfloor & n \leq 5 \\ \lfloor n \times rand(7,8) \rfloor & 6 \leq n \end{cases} \quad (2)$$

Since all nodes are able to obtain the number of competing nodes, they form a cooperative game and nodes with fewer numbers of competitors tune a smaller CW$_{min}$. The best strategy for rest of nodes with more competitors is regulation of a larger CW$_{min}$ for reducing collision probability. One advantage of [6, 7] compared with other games is having no need of information exchange like SNR [5]. With respect to partial swarm optimization, [31] has proposed a game called (G-PSO) for WMNs as follows: in the game, mesh routers operate complicated mechanisms like ARMA and Kalman filters [30] and estimate the game state more exactly and broadcast it across all nodes. Thus, router and nodes form a cooperative game with respect to estimated condition so as to reach global best (g$_{best}$). In G-PSO, each node tunes its contention window according to the following equation [31]:

$$CW_{min} = \begin{cases} CW_{min}^{p} & a \; node \; does \; not \; know \; the \; game \; state \\ CW_{min}^{g} & a \; node \; knows \; the \; game \; state \end{cases} \quad (3)$$

[32] proposed a cooperative game (G-EDCA), in which each node estimates the number of users in each traffic class through listening and counting the data frames with different source addresses from every kinds of traffic and then tunes its equilibrium strategy namely the size of contention window for each traffic class in each stage of transmission as follows. This game is also an incomplete cooperative game [32] :

$$CW_{i,min} = \lfloor n_i \times rand(7,8) \rfloor \quad i = 3, 2, 1, 0 \quad (4)$$

In [33], it is supposed that each node knows conditions of its channel-SNR, number of existing nodes as well as probability distribution of channel conditions for other nodes. Then based on this assumption a CSMA game has been proposed. The game is an opportunistic CSMA game based on user's channel conditions and it is robust to selfish users. Since the proposed game is a complete game; cost, probability distribution, and utility functions are similar for all users. Although cooperative algorithms are more effective and optimal, they need information exchange like PRICE, SNR, etc, to be updated.

Wireless nodes usually are not exactly aware of number of nodes in network and each node can obtain some limited information about channel state (e.g. collided packets, busy or idle state of channel) through listening to channel. In such conditions the best thing a node can do, is to optimize its personal goals. Therefore, for modelling such situation, non-cooperative game models are the best choice. It is shown that MAC protocols can be reverse engineered and studied within the game theory framework [2, 22, 23]. A vector of channel access probability *p* is equilibrium if for the given contention of network; no node is willing to change its transmission probability. In [34, 35] has been shown that(EB) exponential back off protocol can be modelled with a non-cooperative game in which links try to maximize their utility function in the form of reward for successful transmission. Therefore, for EB protocol, non-cooperative game is more suitable than global optimization model and it has been shown that without coordination between nodes, selfish behaviours reduce network performance.

In forward engineering, a cooperative NUM problem has been formulated according to persistence probability of each node and links for the purpose of improving medium access protocol [15, 34, 35]. With a little exchange of contending prices the best coordination for balancing the nature of non-cooperative EB and presentation of a new protocol that moves toward the best converging point can be obtained. Convexity feature is necessary for designing an optimal distributed algorithm and each node has a global convex utility function [15, 34, 35]. In [2, 21, 23, 36] there are examples of how utility functions are designed and also it includes determination of contention criterion through reverse engineering from existing protocols and from desirable performance points (according to throughput and fairness) and through forward engineering in a heuristic manner as the following.

$$U_i(p_i) = p_i - \frac{\theta}{1-a}\hat{c}(1 - p_i)^{1-a} \qquad (5)$$

$$v_i(p_i) := \frac{1}{a_i}(\frac{(a_i-1)b_i}{a_i}\ln(u_i p_i - b_i) - p_i) \qquad (6)$$

In [22] service differentiation has been provided through defining different utility functions according to traffic classes with different weights as a non cooperative game. In most researches including [2, 21, 23, 36] only cellular wireless local networks have been taken into account in which each node hears transmissions of other nodes. Also it is assumed that all nodes are saturated, channel is free from error and packet drop occurs only in a result of collision. In many researches e.g.[21, 37], the payoff functions is defined as below:

$$u_i = U_i(p_i) - p_i q_i(p) \qquad (7)$$

$$q_i(p) = 1 - \prod_{j=N-\{i\}}(1 - p_j) \qquad (8)$$

In fact, the cost functions of these researches are linear functions of multiplication of transmission and collision probabilities. In above equations, $U_i(p_i)$ is utility function, $p_i$ is channel access probability for node $i$, and $q_i$ is the criterion for measurement of contention like collision probability, idle time within channel access, etc. but $U_i(p_i)$ has been defined with different models. Authors of [37] has used game theory as an optimization tool. The proposed game is an iterative multi-stage game non-cooperative game with selfish users and Each node does not need know the total number of nodes. The utility function of the game is defined as the following;

$$U_i(p_i) := \frac{p_i(\ln w_i - \ln p_i + 1)}{\ln \gamma_i}, \quad \gamma_i = \frac{w_i}{v_i}, \quad v_i \le p_i \le w_i \qquad (9)$$

[38]studies behavior of non-cooperative users who tune their access probability through changing their persistence coefficient or the back off exponential control parameter. It has defined the disutility function for each user according to collision rate and power consumption. In [38] like other researches, the effect of physical layer and noise has not been considered and frame loss has been regarded only because of collision.

In studies like [39-42] channel contention problem has been implemented as a non-cooperative power control game (GMAC) in a distributed manner with quasi CSMA/CA access mechanism. GMAC uses a shared channel for data and control. In definition of the utility, GMAC uses a linear pricing factor of power consumption.

$$u_i(p_i, p_{-i}) = \log(1 + \gamma_i) - \alpha_i p_i \qquad (10)$$

In the game, $\gamma_t$ denotes the SINR at the receiver and power control decisions are made for each packet and each receiver uses the knowledge of previous scheduled transmissions among its neighbors in order to join the scheduling program. Simulations have shown that GMAC makes possible more concurrent transmissions and improves power consumption.

In [43, 44], using static non-cooperative game a distributed power-aware MAC algorithm called PAMG has been modeled for Ad Hoc networks. In these games, each active link has been taken as a player. Strategies vector is two-dimensional including transmission probability and power probability. In the game, each node according to its place in the network and received feedback from the channel, tune their transmission probability and power selfishly. Decision making in each link occurs based on optimality of a payoff function that has been defined on two-dimensional strategy space and the value of this function shows the payoff resulted by medium. Each link can improve its data rate through increasing its power and transmission probability. Authors of [45] has proposed a non-cooperative and contention-based medium access game (CAG) with initial frameworks like [22] and with selfish users. In CAG the number of users is pre-determined and constant. The strategy of each user is his transmission probability $\alpha_i$. Payoff function has been also defined through combination of payoffs resulted by successful transmission $u_s$, collision $u_c$, and waiting $u_w$ as follows:

$$u(\alpha_i) = u_s - u_c + u_w = A\alpha_i \prod_{j \in N(i)}(1 - \alpha_j) - B\alpha_i\left(1 - \prod_{j \in N(i)}(1 - \alpha_j)\right) + C(1 - \alpha_i) \quad (11)$$

Then CAG has converted into an optimization problem with Constraint condition and the strategy has been updated by gradient for reaching Nash equilibrium. Author of [46] has considered a scenario same as [47] with presence of a specific number of users with different QoS requirements. She modelled the priority of traffic in this CSMA/CA network through game theory. Each node labels its traffic with a specific traffic class and these classes do not change. In this mode, it is expected that nodes with higher priorities take over the wireless resources and access the wireless medium with lower values of contention window. Nodes with low priority should not continue reducing their contention window to eliminate the higher priority nodes action in order to have more medium access. Each time when throughput of a node is estimated, punishment mechanism through jamming of those nodes which have violated the expected throughput, does not allow for confirmation of proper reception of the traffic transmitted by a cheater to a specific destination and thus the transmitter does not receive any acknowledgement [46]. Yang formulated a non-cooperative game [48, 49], in which each link tries selfishly to maximize its payoff function, its throughput, by its local information. Power constraints have been also added to the conditions and price of each link reflects the interference it gets from the price receiver. Although performance of non-cooperative game does not equate cooperative game but in case of using non-cooperative games, message exchange cost reduces. [3] Has proposed a non-cooperative persistence model with purpose of minimizing the exchanged information between nodes. In the non-cooperative game, users access the channel in order to maximize their successful transmissions based on collision cost and power consumption. In [50] a two-stage mathematical model called TRG/CSMA has been proposed. in the first step, throughput and delay have been selected as optimization goals and a game between (n) nodes is played. In the second step of game, throughput and delay as players play a two-players game with infinite strategy and obtain their weight in the game and contention window is tuned.

Study of cooperative and non-cooperative models proposed so far for the purpose of improving carrier sense multiple access model leads toward this fact that the desirable solution for carrier sense multiple access games should be unique, fair, and Pareto optimal. In cooperative games, using bargaining techniques, etc, these conditions can be obtained. In non-cooperative games, however, these techniques are not suitable and to reach the desirable result, some other techniques like punishment should also be used [17].

## 5. CONCLUSIONS

In recent years application of game theory has been introduced in wireless communications. Applications of game theory in data link layer is related to medium access control problem and in these games, selfish users through unfair sharing in channel access try to maximize their performance. This reduces the ability of other users in accessing the channel. Users have to overcome the collision incurred by limited transmission resources, thus game theory has turned into a powerful tool to analyze and improve the contention-based protocols. Numerous games have been proposed for modelling such environment in which behavior of users is (transmit and wait) and in some others behaviours include transmission probability. Studying the researches showed that cooperative random access can use radio resources more effectively and fairly compared to non-cooperative random game. Thus for achieving an efficient and fair system with low collision, reception of controlling messages from other nodes seems necessary. Of course, it should be noted that message passing between nodes is based on cooperation assumption with purpose of acquiring a system with a global objective.

## ACKNOWLEDGMENT

This work is partially supported by Iran Telecommunication Research Canter (ITRC No.18507/500T).

**Authors**

**Mahdieh Ghazvini** received her B.Sc. from Shahid Bahonar University, Kerman, Iran in 2000, and her M.S. from the University of Isfahan, Isfahan, Iran in 2004 in Computer Architecture Engineering. Currently she is a PhD. candidate of Computer Architecture Engineering at the University of Isfahan. Her research interests are wireless networks, game theory, signal processing and neural networks.

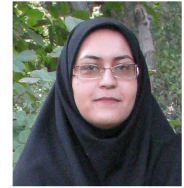

**Naser Movahedinia** received his B.Sc. from Tehran University, Tehran, Iran in 1987, and his M.Sc. from Isfahan University of Technology, Isfahan, Iran in 1990 in Electrical and Communication Engineering. He got his PhD. degree from Carleton University, Ottawa, Canada in 1997, where he was a research associate at System and Computer Engineering Department, Carleton University for a short period after graduation. Currently he is an associate professor at the Computer Department, University of Isfahan. His research interests are wireless networks, signal processing in communications and Internet Technology .

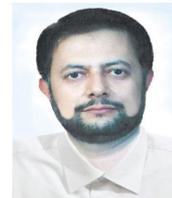

**Kamal Jamshidi** received his B.Sc. from Isfahan University of Technology, Isfahan, Iran in 1988, and his M.Sc. from Isfahan University of Technology, Isfahan, Iran in 1991 in Control and Instrumentation Engineering. He got his Ph.D. degree from IUT University, India in 1995, Fuzzy control. Currently he is an assistant professor at the Computer Department, University of Isfahan. His research interests are wireless networks, digital control, and fuzzy logic.

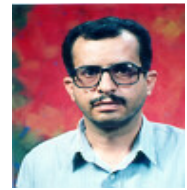